\documentclass[twocolumn,amsmath,amssymb,prb,longbibliography]{revtex4-2}
\usepackage{xcolor}
\usepackage{multirow}
\usepackage{array}
\usepackage{amsmath}
\usepackage{graphicx}% Include figure files
\usepackage{dcolumn}% Align table columns on decimal point
\usepackage{bm}% bold math
\usepackage{verbatim}
\DeclareMathAlphabet \mathbfcal{OMS}{cmsy}{b}{n}

\begin{document}

\title{Selective enhancement of Coulomb interactions in planar Weyl fermions}
\author{ Vadym Apalkov$^a$}
\author{Wenchen Luo$^b$}
\author{ Tapash Chakraborty$^c$}

\affiliation{$^a$Department of Physics and Astronomy, Georgia State University, Atlanta, Georgia,30303, USA, $^b$ School of Physics, Central South University, Changsha, Hunan 410083, China, $^c$ Department of Physics and Astronomy, University of Manitoba, Winnipeg, MB, Canada}

\date{\today}

\begin{abstract}
We report on our study of the electron interaction effects in topological two-dimensional (2D) materials placed in a quantizing magnetic field. Taking our cue from a recent experimental report, we consider a particular case of bismuthene monolayer with a strong spin-orbit interaction which can be a Weyl semimetal when placed on a specially tuned substrate. Interestingly, we observe that in some Landau levels of this material, the interaction effects are enhanced compared to those for a conventional 2D system and graphene monolayer. Such an enhancement of electron-electron interactions in these materials is largely due to an anisotropy present in the materials. Additionally, the interaction effects can be tuned by changing the coupling to the substrate and the strongest inter-electron interactions are observed when the system is a Weyl semimetal. The observed enhancement of the interaction effects can  therefore be an important signature of the 2D Weyl fermions.  
\end{abstract}

\maketitle

Discovery of the integer quantum Hall effect (IQHE) in 1980, which originated in uniform two-dimensional
(2D) electron gas in a perpendicular high magnetic field \cite{40_years_QHE}, ushered in a new era of condensed matter
physics \cite{Encyclopedia}. Subsequently, it was realized that this effect can also in principle be associated with a topological 
invariant of 2D band structure, the Chern number \cite{Quantized_Hall_Conductance_Two_Dimensional_Periodic_Potential}, that can be interpreted in terms of the Berry curvature \cite{Holonomy}.
Soon after the discovery of IQHE, fractional quantum Hall effect (FQHE) was discovered \cite{Two_Dimensional_Magnetotransport,Two_dimensional_electron_correlation}, and the
origin of that remarkable effect was explained as a direct result of strong electron-electron interactions \cite{Anomalous_Quantum_Hall_Effect}. 
Those interactions lead to a unique incompressible state\cite{Book_FQHE}, where the fractionally-charged low-energy
quasiparticles obey exotic fractional statistics\cite{Encyclopedia,Fractional_Statistics,Anyon_collision} . The incompressible state was also found to display many other important physical properties 
\cite{Book_FQHE,Spin_reversed_ground_state,
FQHE_tilted_field,
Evidence_phase_transition_FQHE,Spin_configurations_FQHE}, and exploration of interacting 2D electron gas dominated research
in condensed matter physics. However, research on electronic states took a dramatic turn after successful 
fabrication of graphene\cite{Encyclopedia,graphene_advances_2010} and topological insulators \cite{Ando_book_E} when the low-energy dynamics of 2D electrons were found 
to obey the Dirac equation, rather than the usual Schr\"odinger equation and soon the concepts of topology 
and chirality driven electronic properties came to the fore. The age of exploration of quantum materials then began
in earnest \cite{topology_electron_band_sstructure}. 
Although the endless array of unique physical properties has now enriched the field of condensed matter \cite{Interplay_electronic_order,quantum_anomalous_hall}, the topological quantum materials are also expected to contribute significantly in the emerging field of topological electronics\cite{Topological_electronics}.
For interacting Dirac fermions, many-body interaction-driven properties, such as the FQHE, were 
found to be present in monolayer, bilayer and double-layer graphene systems 
\cite{FQHE_graphene,
phase_transitions_bilayer,
stable_pfaffian_bilayer,
Interacting_fermions_book,
Traits_characteristics_interacting_Dirac_fermions,
Interlayer_FQHE}, and in topological insulators \cite{Interacting_Dirac_Fermions_topological_insulator}. 
The next major advancement in search of topological materials  was finding the Weyl fermions, the zero-mass solution of the Dirac equation,  in condensed matter systems \cite{Topological_properties_Dira_Weyl} and the properties of   topological 
semimetals \cite{Quasiparticle_interference_Weyl,
recent_progress_in_staudy_topological_semimetals,Dirac_semimetals_PRL_2015} received serious attention. Here also a new milestone was achieved in 2024 when the
2D Weyl semimetals were reported \cite{Realization_two_dimensional_Weyl_semimetal}. 

Electronic states of topological quantum materials in 2D have the clear advantage over their three-dimensional counterparts because they are easily manipulated by the externally applied fields. While research in quantum materials has been largely based on the non-trivial topological properties of the electronic bands, a microscopic approach to the interaction-induced many-body properties is important to fully understand the intricacies of these novel materials. The advantages of having Weyl semimetals in reduced dimensions 
are manifold: in addition to their new physical characteristics, they can be easily integrated in devices of low-dimensions, 
and the theoretical techniques developed for decades to investigate other low-dimensional systems can possibly be applied 
here as well (with suitable modifications). 

In this work, we report on the quantizing magnetic field induced properties of interacting 
Weyl fermions in two dimensions. We found several unconventional features in the system. More specifically, we analyze a special case of bismuthene monolayer\cite{Realization_two_dimensional_Weyl_semimetal}, which in a free-standing case behaves as gapped anisotropic graphene-like material, where the gap is determined by the spin-orbit interaction, 
but on a specially chosen substrate, e.g., the SnSe substrate, it becomes a Weyl semimetal. When such a system is placed in a quantizing magnetic field, we find several remarkable features: an enhancement of electron-electron interactions within a single Landau level compared to a conventional two-dimensional materials. 
This enhancement is largely due to the anisotropy present in the system, which results in the system's Landau levels that are the mixtures of the conventional 2D Landau levels. When the bismuthene monolayer is placed on a substrate, which breaks the inversion symmetry and produce a Weyl semimetal with two valleys, the electron-electron interactions become even more enhanced in one of the valleys of the system.

As mentioned above, the system of two-dimensional Weyl semimetal adopted here has been realized in bismuthene monolayer placed on a SnSe substrate\cite{Realization_two_dimensional_Weyl_semimetal}. The substrate brakes the inversion symmetry of the system and results in the formation of two Weyl points with low-energy linear dispersion. Near the Weyl points the low-energy effective electron Hamiltonian
 takes the form 
\begin{equation}
{\cal H}_{\tau} = 
\tau \left(v_x k_x \sigma_x  + 
\Delta k_x \right)  + v_y k_y \sigma_y 
+ \tau \lambda_{SOC} \sigma_z s_z + \lambda_S \sigma_z,
\label{H0}
\end{equation}
where $\sigma_i$ and $s_i$ are Pauli matices corresponding to sublattice and spin degrees of freedoms, respectively. The system has the following parameters: Fermi velocities along the $x$ and $y$ directions, $v_x = 3.17\times 10^5$ m/s, $v_y = 4.23\times 10^5$ m/s, tilting of the Weyl cone in the $k_x$ direction, $\Delta = 0.19 \times 10^5$ m/s, and the spin-orbit coupling, $\lambda_{SOC} = 55$ meV. 
Coupling of bismuthene monolayer to the substrate introduces the term 
$\lambda_S \sigma_z$, which breaks the inversion symmetry and if $\lambda_S = \lambda_{SOC}$ then the energy dispersion is gapless with two Weyl points which are labeled by the parameter $\tau =\pm 1$. In what follows, we consider $\lambda_S$ as a variable parameter of the system, which can be controlled by using different substrates. The Hamiltonian (\ref{H0}) is similar to the graphene Hamiltonian with two fundamental differences introduced by the anisotropy terms. The anisotropy terms in (\ref{H0}) are of two types:  the anisotropy due to difference between the Fermi velocities, $v_x$ and $v_y$, and the anisotropy due to the parameter $\Delta$, which characterizes the tilting of the Weyl cones.  The important role that these anisotropies play in our system will soon be clear below.

In the presence of a magnetic field $B$ perpendicular to the bismuthene monolayer, the electron momentum is replaced with $\mathbf{\pi} = \mathbf{p} - e\mathbf{A}$, where $\mathbf{A}$ is the corresponding vector potential. In the symmetric guage, the vector potential is $\mathbf{A} = (-By/2, Bx/2,0)$.  
Then the Landau levels can be calculated by expressing the corresponding wave functions in the bases of Landau functions $\phi_{n,m}$ of the conventional electron system with parabolic energy dispersion, where $n$ is the corresponding Landau level index and $m$ is the $z$ component of the angular momentum. 
Below we consider only one component of the electron spin, say $s_z=-1$, assuming that the levels of the system with different spin components are separated by the Zeeman energy. Therefore, for a given spin component, the Landau level wavefunctions can be expressed as 
\begin{equation}
\Psi_{n,m} = \left(
\begin{array}{c}
\sum_{n_1} C_{n,n_1} \phi_{n_1,m} \\
\sum_{n_1} D_{n,n_1} \phi_{n_1,m}
\end{array}
\right),
\label{wave1}
\end{equation}
where $C_{n,n_1}$ and $D_{n,n_1}$ are unknown coefficients which can be found from solutions of the corresponding eigenvalue equation ${\cal H}_{\tau} \Psi_{n,m} = E_n \Psi_{n,m}$ and $E_n$ is the Landau energy spectrum. In the expansion, Eq.\ (\ref{wave1}), of the wavefunctions $\Psi_{n,m}$ in terms of 
convensional Landau level functions $\phi_{n,m}$ we consider a finite 
number of functions $\phi_{n,m}$, $n=0,\ldots, N$. Then, for a given Weyl point $\tau=\pm 1$, the size of the corresponding Hamiltonian matrix ${\cal H}_{\tau}$ is $2(N+1)$. From diagonalization of such a matrix, we obtain the Landau energy spectrum, $E_n$, and the coefficients 
$C_{n,n_1}$ and $D_{n,n_1}$. In our studies below, we consider $N=100$.

The two types of anisotropy in Hamiltonian (\ref{H0}), i.e., tilting term $\Delta $ and different Fermi velocities along the $x$ and $y$ directions, result in different effects on the Landau level wavefunctions (\ref{wave1}). Even for an isotropic case of graphene, the isotropic Fermi velocity results in the mixing of conventional Landau levels belonging to different sublattices, e.g., the Landau wavefunctions for graphene have the following structure $(\phi_{n-1,m}, \phi_{n,m})$ for $n\geq 1$. In this case of bismuthene monolayer, the anisotropy in Fermi velocities results in additional mixing of conventional Landau wavefunctions belonging to different sublattices. At the same time, the presence of the tilting term $\Delta$ results in the mixing of conventional Landau wavefunctions belonging to the same sublattice.

The electron-electron interaction within a single Landau level is characterized in terms of the Haldane pseudopotentials, $V_m$, which are the energies of the interaction of two electrons with relative angular momentum $m$ \cite{Haldane_Rezayi_1985}. With the known Landau level wavefunctions (\ref{wave1}), the Haldane pseudopotentials have the following form
%\begin{eqnarray}
%& &  V_m^{(n)} = \sum_{n_1, n_2, n_3, n_4} \left( C_{n,n_1}^* C_{n,n_4} +  
%D_{n,n_1}^* D_{n,n_4} \right)    \nonumber \\
%& & \times \left( C_{n,n_2}^* C_{n,n_3} +  
%D_{n,n_2}^* D_{n,n_3} \right)  \delta_{n_1-n_4+n_2-n_3} 
% \nonumber \\ 
%& & 
%\times \sqrt{\frac{m_{1,4}!m_{2,3}!}{M_{1,4}!M_{2,3}!}} %\int_0^{\infty } \frac{dq}{2\pi} qV(q) \left(\frac{q^2}{2}\right)^{|n_1-n_4|}  
%   \nonumber \\
%& & \times L_{m_{1,4}}^{|n_1-n_4|}\left(\frac{q^2}{2}\right) 
% L_{m_{2,3}}^{|n_1-n_4|}\left(\frac{q^2}{2}\right)  L_m %(q^2) e^{-q^2},
%\label{Vmm}
%\end{eqnarray}
\begin{eqnarray}
V_{m}^{\left( n\right) }= & & \sum_{n_{1},\ldots n_{4}=0}^{N}\delta
_{n_{1}-n_{4}+n_{2}-n_{3}}    \int   \frac{qdq}{2\pi }V\left( q\right)
e^{-q^{2}} 
\nonumber  \\
& & \times L_{m}\left( q^{2}\right)  
        F_{n_{1},n_{4}}\left( q\right) 
F_{n_{2},n_{3}}\left( q\right) , 
\label{Vmm} 
\end{eqnarray}
\begin{eqnarray}
 F_{n_{1},n_{4}}\left( q\right)  =& &\left( C_{n,n_{1}}^{\ast
}C_{n,n_{4}}+D_{n,n_{1}}^{\ast }D_{n,n_{4}}\right)  
\nonumber \\
& & \times \sqrt{\frac{m_{1,4}!}
{M_{1,4}!}}\left( \frac{q}{\sqrt{2}}\right) ^{\left\vert
n_{1}-n_{4}\right\vert }L_{m_{1,4}}^{\left\vert n_{1}-n_{4}\right\vert
}\left( \frac{q^{2}}{2}\right)  \nonumber 
\end{eqnarray}
where $V(q) =  \frac{e^2}{\epsilon l_B q }$ is the Coulomb interaction in the momentum space, $l_B$ is the magnetic length, $e$ is the electron charge, $\epsilon$ is the dielectric constant,  $L_n(x)$ are the Laguerre polynomials, $L_n^{n_0}(x)$ are associated Laguerre polynomials, $m_{i,j} = \min{(n_i,n_j)}$, and $M_{i,j} = \max{(n_i,n_j)}$. The Haldane pseudopotentials defined by Eq. (\ref{Vmm}) are based on the projection of the electron change density on a given Landau level\cite{FQHE_graphene,MacDonal_PRL_2006,Goerbig_interaction_graphene_2006,
composite_fermion_graphene_2007}. Another expression for the Haldane pseudopotentials \cite{macdonald_PRL_2010_graphene,density_states_graphene_2011} considers the coupling between sublattice and orbital degrees of freedom and expresses the pseudopotentials in term of eigenfunctions of the two-particle system. Compared to Eq. (\ref{Vmm}), this type of Haldane pseudopotentials have smaller values at odd relative angular momenta $m$.  

\begin{figure}[b]
\includegraphics[width=\columnwidth]{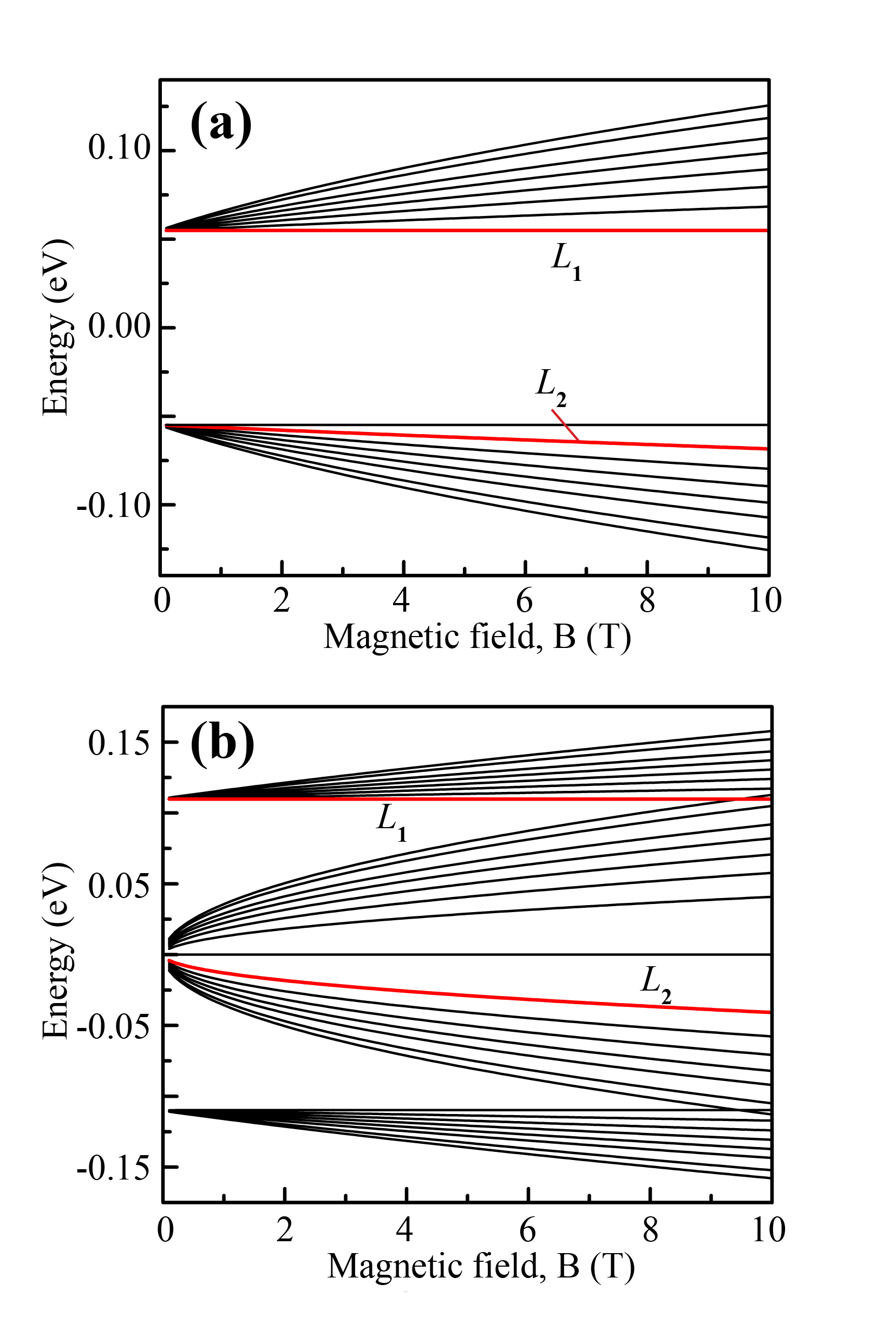}
\caption{\label{fig_LL}   Landau levels of a bismuthene monolayer with two values of substrate coupling, $\lambda_S$: (a) $\lambda_S= 0$, which corresponds to free-standing monolayer, (b) $\lambda_S = \lambda_{SOC}=55$ meV, which corresponds to the Weyl semimetal case. The results are shown for only one component of the electron spin, $s_z = -1$. The red lines show the Landau levels with strong electron-electron coupling. Such a coupling is characterized by the corresponding Haldane pseudopotentials. The Landau levels with strong electron-electron coupling are labeled as $L_1$ and $L_2$ levels. Two sets of Landau levels correspond to two Weyl points, $\tau = \pm 1$, where the set of Landau levels, which is originated from the spectrum with zero bandgap, corresponds to the Weyl point $\tau = 1$, while another set, with a finite bandgap, corresponds to the Weyl point $\tau = -1$.} 
\end{figure}

The Landau levels of the bismuthene monolayer are shown in Fig.\ \ref{fig_LL} for two values of $\lambda_S$, corresponding to the case of free standing bismuthene monolayer, $\lambda_S = 0$ [Fig.\ \ref{fig_LL}(a)] and a monolayer on a substrate with $\lambda_S = \lambda_{SOC}$ 
[Fig.\ \ref{fig_LL}(b)]. The results are shown for two Weyl points, i.e., for $\tau = \pm 1$. For  $\lambda_S = 0$ and without the magnetic field the system has a bandgap of $2\lambda_{SOC}$, which is also clearly visible in the corresponding Landau spectrum. For $\lambda_S=\lambda_{SOC}$, the gap is closed and the Weyl semimetal is realized. In this case, for one of the Weyl points, $\tau = 1$, i.e., for one of the valleys, the Landau spectrum shows the gapless behavior with one of the levels having zero energy, while for the other valley, $\tau =-1$, the Landau energy spectrum shows a finite gap with the value of $4\lambda_{SOC}$. In both panels, the Landau levels with the strongest electron-electron interactions, i.e., the largest values of the corresponding Haldane pseudopotentials, are shown by red lines. The levels are marked as $L_1$ and $L_2$ [Fig.\ \ref{fig_LL}].

%%%%%%%%%
%1%and Fig.~\ref{fig1}.%

%%%%%

\begin{figure}[b]
\includegraphics[width=\columnwidth]{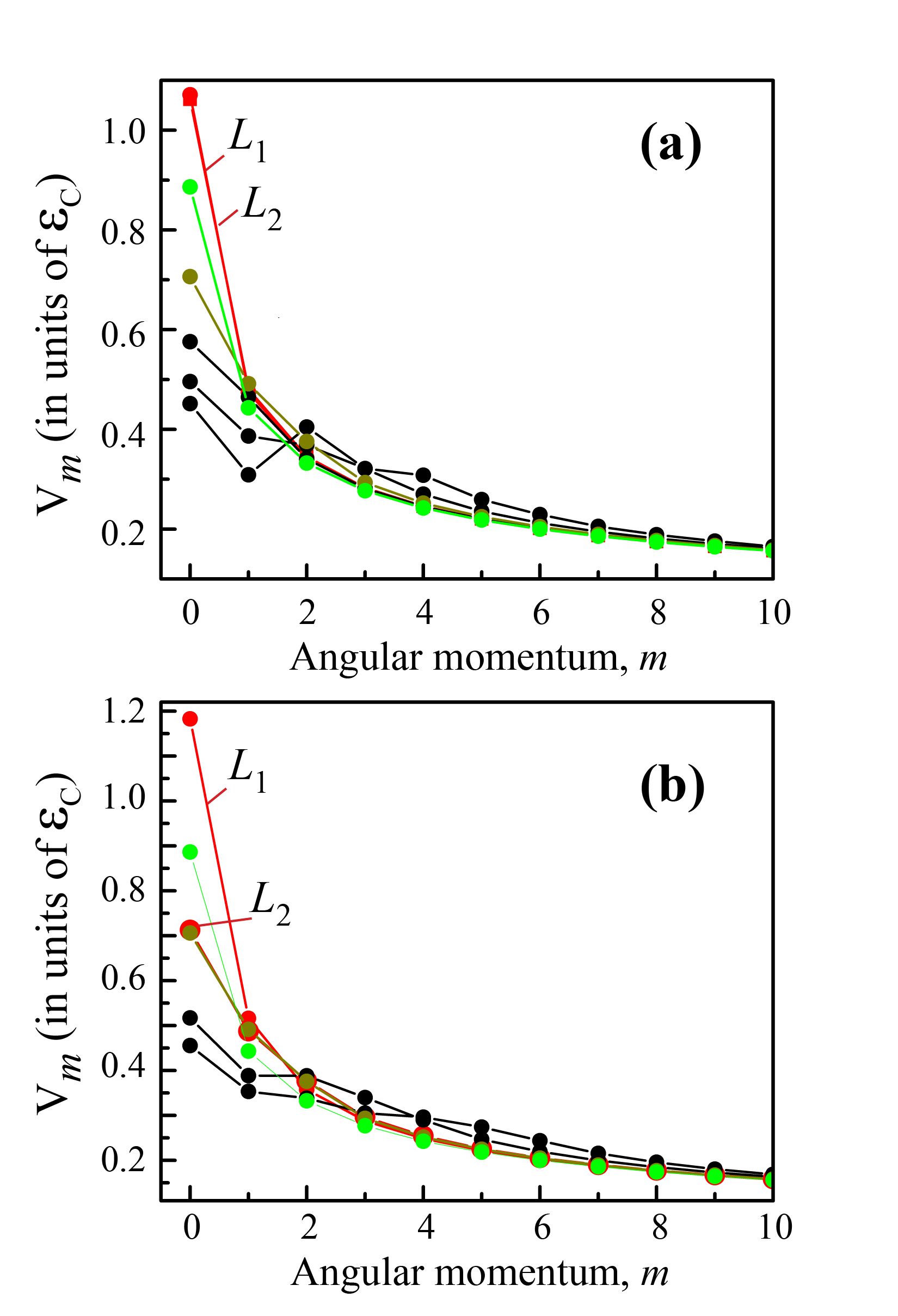}
\caption{\label{fig_Vm}   Haldane pseudopotentias at a few Landau levels of a bismuthene monolayer with two values of substrate coupling, $\lambda_S$: (a) $\lambda_S= 0$, (b) $\lambda_S = \lambda_{SOC}=55$ meV, which corresponds to the Weyl semimetal case. The largest Haldane pseudopotentials are realized at Landau levels $L_1$ and $L_2$ and are shown by red dots. The dark green dots  show the Haldane pseudopotentials for a conventional  electron system with parabolic energy dispersion at the $n=0$ Landau level, while the dark orange dots show the Haldane pseudopotentials for a graphene system at the $n=1$ Landau level.    } 
\end{figure}

The Haldane pseudopotentials, $V_m$, for a few Landau levels of the system are shown in Fig.\ \ref{fig_Vm}. The largest pseudopotentials are realized at the Landau levels $L_1$ and $L_2$, which are marked by red dots in Fig.\ \ref{fig_Vm}. We also show the Haldane pseudopotentials for a conventional electron system with parabolic energy dispersion  for the Landau level $n=0$ by dark green dots and for graphene monolayer for the Landau level $n=1$\cite{FQHE_graphene} by dark orange dots.  The data in Fig.\ \ref{fig_Vm} show that electron-electron interactions are enhanced for a bismuthene monolayer compared to the conventional case and graphene with the strongest enhancement is visible for small values of $m$. Also, the strongest electron-electron interactions are realized for the Landau level $L_1$ for the case of the Weyl semimetal [Fig.\ \ref{fig_Vm}(b)] when $\lambda_S = \lambda_{SOC}$. Under this condition, 
for the Landau level $L_2$, the Haldane pseudopotentials are almost the same as the ones in graphene. 

\begin{figure}[b]
\includegraphics[width=\columnwidth]{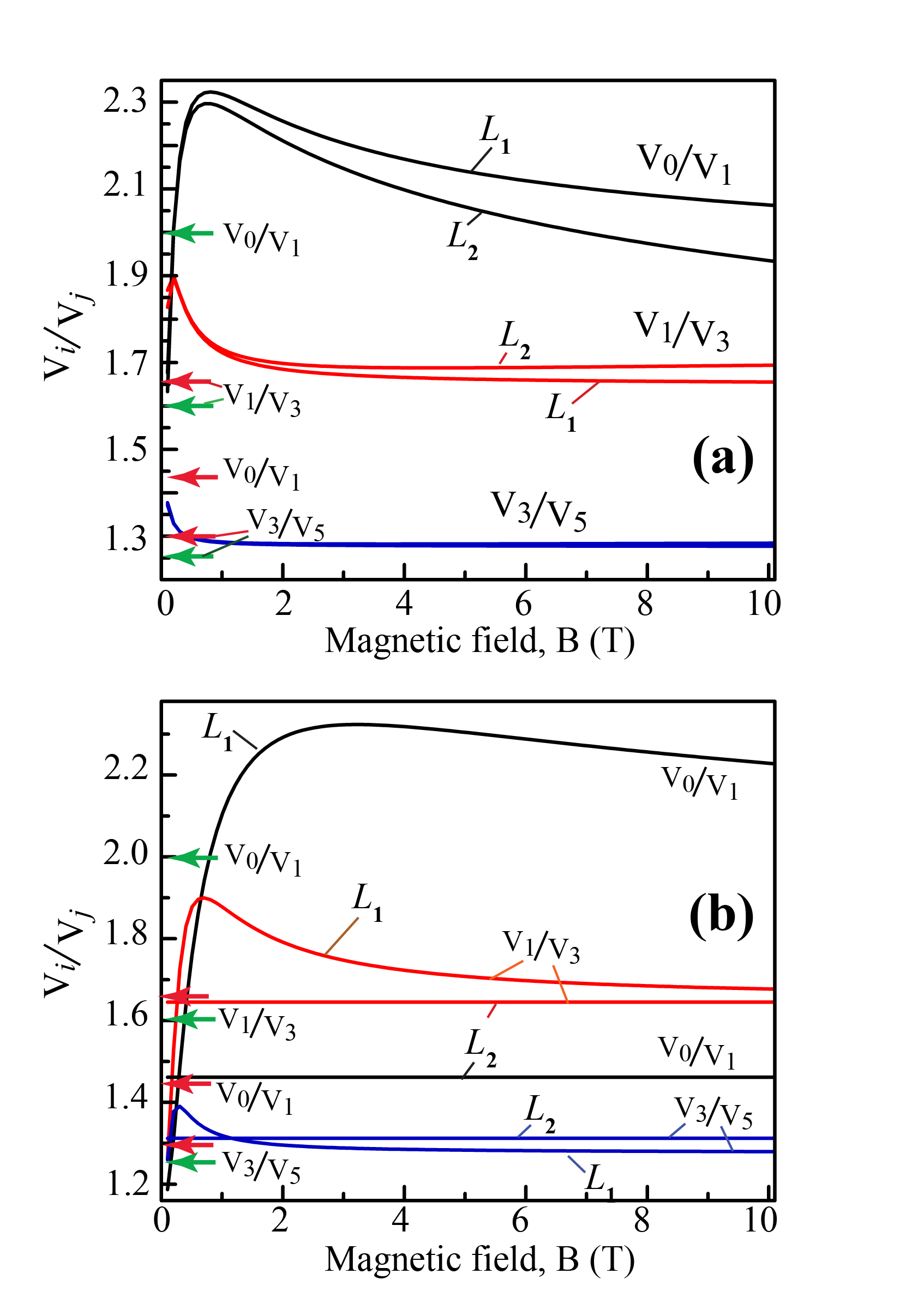}
\caption{\label{fig_Vm_vs_B}  Ratios of Haldane pseudopotentias $V_0/V_1$, $V_1/V_3$, and $V_3/V_5$ at $L_1$ and $L_2$ Landau levels as a function of the magnetic field are shown for a bismuthene monolayer. The value of the substrate coupling, $\lambda_S$, is $\lambda_S= 0$ (a) and $\lambda_S = \lambda_{SOC}=55$ meV (b). In panel (a), the results for both $L_1$ and $L_2$ Landau levels are the same, while in panel (b), the data for Landau levels $L_1$ and $L_2$ are marked in the picture. Green arrows  show the results for the case of a conventional  electron system with the parabolic energy dispersion at $n=0$ Landau level, and the red arrows show the results for the case of a graphene monolayer at $n=1$ Landau level. For the case of a Weyl semimetal shown in panel (b), the ratios of pseudopotentials  $V_0/V_1$, $V_1/V_3$, and $V_3/V_5$ for the Landau level $L_2$ do not depend on the magnetic field and are almost the same as the corresponding ratios for graphene. 
 } 
\end{figure}

One of the characteristic features  of electron-electron interactions is how fast the corresponding Haldane pseudopotentials decay with increasing  relative angular momentum, $m$. Such a decay determines the stability of the FQHE states and is characterized by the ratio of Haldane pseudopotentials, $V_i/V_j$, for example, $V_0/V_1$, $V_1/V_3$, and $V_3/V_5$. These ratios are shown in Fig.\ \ref{fig_Vm_vs_B} as a function of the magnetic field. For a free-standing bismuthene monolayer [Fig.\ \ref{fig_Vm_vs_B}(a)] the results for  $L_1$ and $L_2$ Landau levels are close with small deviation at large magnetic fields. The ratio $V_0/V_1$ shows a nonmonotonic dependence on the magnetic field with the maximum value realized at magnetic field of around $1$ T, while both $V_1/V_3$ and $V_3/V_5$ are saturated at a magnetic field of around 1 T and remain constant for $B>1$ T. The corresponding ratios of 
the pseudopotentials for the conventional case at Landau level $n=0$ and graphene at Landau level $n=1$ are shown by green and red arrow, respectively. The results in Fig.\ \ref{fig_Vm_vs_B}(a) clearly show that the ratios $V_0/V_1$ and $V_1/V_3$ are strongly enhanced for a free standing bismuthene monolayer compared to both the conventional system and graphene, especially for small values of the magnetic field, $B<2$ T.

In the case of the Weyl semimetal [Fig.\ \ref{fig_Vm_vs_B}(b)] which corresponds to the condition of $\lambda_S = \lambda_{SOC}$, the results for the Landau levels $L_1$ and $L_2$ are different and the corresponding ratios show very different behaviors. For the $L_2$ Landau level, the pseudopotentials do not depend on the magnetic field due to effective cancellation of a spin-orbit term in the Hamiltonian (\ref{H0}) by the substrate coupling, i.e., $\lambda_S-\lambda_{SOC} = 0$ for $\tau = 1$. In this case, the ratios of pseudopotentials for the $L_2$ level are almost the same as the corresponding ratios for the graphene monolayer at the Landau level $n=1$, which are shown by red arrows. 
For the $L_1$ Landau level, all ratios show nonmonotonic dependence on $B$ with clear maxima realized at small magnetic fields. Also the ratio $V_0/V_1$ for the $L_1$ Landau level is much larger than the corresponding ratio for $L_2$ Landau level, while the ratios $V_1/V_3$ and $V_3/V_5$ for both Landau levels are almost the same for large magnetic field, $B> 6$ T. For the $L_1$ Landau level, all the ratios are much larger than the corresponding ratios for the conventional case and graphene, especially for small magnetic field less than 3 T. 

% the ratios $V_1/V_3$ and $V_3/V_5$ as a function of the magnetic field are constant for $B>2$ T, but $V_0/V_1$ monotonically increases with $B$. While  $V_3/V_5$ for both Landau levels, $L_1$ and $L_2$, are almost the same, the ratio $V_1/V_3$ for the Landau level $L_1$ is much larger than the one for the level $L_2$. 

\begin{figure}[b]
\includegraphics[width=\columnwidth]{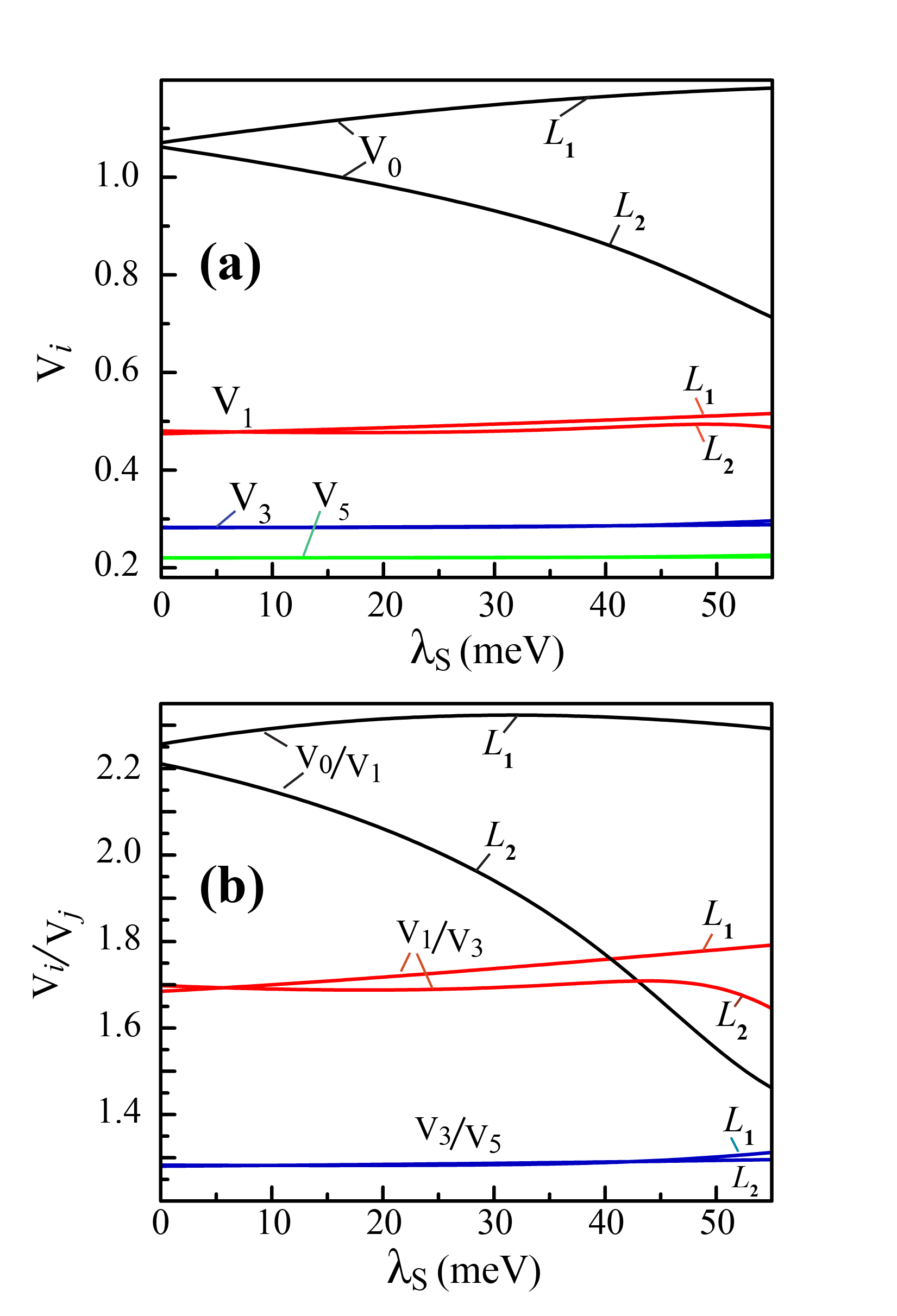}
\caption{\label{fig_Vm_vs_lambda}   Haldane pseudopotentias as a function of the substrate coupling, $\lambda_S$. The pseudopotentias are shown for two Landau levels, $L_1$ and $L_2$. (a) The Haldane pseudopotentials $V_0$, $V_1$, $V_3$, and $V_3$ are shown as a function of $\lambda_S$. (b) The ratios of  the pseudopotentials, $V_0/V_1$, $V_1/V_3$, and $V_3/V_5$ are shown as a function of $\lambda_S$. Here, $\lambda_S=0$ corresponds to a free-standing bismuthene monolayer, while $\lambda_S = 0.055$ eV corresponds to a Weyl semimetal case. 
The applied magnetic field is $B=2$ T.  
 } 
\end{figure}

To examine the properties of the system as the coupling of a bismuthene monolayer to the substrate changes from the free-standing case to the Weyl semimetal, we plot the first few Haldane pseudopotentials and the corresponding ratios as a function of $\lambda_S$ in Fig.\ \ref{fig_Vm_vs_lambda}. The results are shown for both $L_1$ and $L_2$ Landau levels. In Fig.\ \ref{fig_Vm_vs_lambda}(a), the pseudopotentials 
$V_0$, $V_1$, $V_3$, and $V_5$ are presented. For $\lambda_S = 0$, the values of these pseudopotentials are almost the same for $L_1$ and $L_2$ Landau levels. With increasing $\lambda_S$, the pseudopotentials monotonically increase and decrease for levels $L_1$ and $L_2$, respectively. The difference between the levels $L_1$ and $L_2$ is more pronounced for the pseudopotential $V_0$, which is much larger for the $L_1$ level. Pseudopotential $V_1$ also shows a small enhancement for the $L_1$ level compared to the $L_2$ level, while pseudopotentials $V_3$ and $V_5$ are almost the same for both levels $L_1$ and $L_2$ for all values of $\lambda_S$. 
 The behavior of pseudopotentials shown in Fig.\ \ref{fig_Vm_vs_lambda}(a) suggest that with increasing $\lambda_S$, i.e., by increasing the coupling to the substrate, the electron-electron interactions are enhanced for the Landau level $L_1$ but are suppressed for the Landau level $L_2$. One of the effect of the substrate is suppression of the band gap, i.e., with increasing $\lambda_S$ the band gap decreases and for $\lambda_S=\lambda_{SOC}$ the band gap becomes zero. Therefore, with decreasing  band gap of the  
 bismuthene monolayer the electron-electron interactions are enhanced for the level $L_1$ and suppressed for the level $L_2$. 
 
 The suppression of the electron-electron interactions in gapped 
 graphene monolayer with increasing the band gap was also reported in Ref.\ \cite{mass_term_graphene_2016}. 
 To further  illustrate the dependence on the interactions  on $\lambda_S$  we evaluate the $\nu=1/3$-FQHE energy gaps, $\Delta_{1/3}$\cite{Book_FQHE}, for different values of $\lambda_S$. The calculations were done in the spherical geometry\cite{Haldane_Rezayi_1985} with seven electrons in the system and the magnetic field of 2 T. When $\lambda_S=0$, for both $L_1$ and $L_2$ levels, the gap is around $\Delta_{1/3}\approx 0.095 \epsilon_C$, where $\epsilon_C = e^2/\epsilon l_B$  is the Coulomb energy. With increasing $\lambda_S$, the gap $\Delta_{1/3}$ monotonically increases for the $L_1$ level and reaches the value of 0.11 $\epsilon_C$ at $\lambda_S = \lambda_{SOC}$. For the $L_2$ level, the gap $\Delta_{1/3}$ monotonically decreases with $\lambda_S$ and becomes 0.088 $\epsilon_C$ at $\lambda_S = \lambda_{SOC}$. It is worth noting that in all the cases, the gaps are larger than that of the 1/3-FQHE 
in conventional semiconductors or graphene which is about 0.07 $\epsilon_C$\cite{FQHE_graphene}. 
 The Weyl semimetal is therefore an ideal incompressible Laughlin state with large excitation gaps that could perhaps be experimentally observed.

The ratios of the pseudopotentials, $V_m$, are also shown in Fig.\ \ref{fig_Vm_vs_lambda}(b) as a function of $\lambda_S$. For the level $L_1$, the ratios remain almost constant within the whole range of $\lambda_S$, 
while for the level $L_2$, the ratio $V_0/V_1$ monotonically decreases with $\lambda_S$ from 2.2 value at $\lambda_S =0$ to its 1.45 value at $\lambda_S = \lambda_{SOC}$. For the $L_2$ level, the ratio $V_1/V_3$ remains almost constant with a small suppression at $\lambda_S$ close to 
$\lambda _{SOC}$. 
 The selective enhancement of interaction effects is therefore manifestly clear from these results.

We  note here that the electron-electron interaction effects in three-dimensional (3D) Weyl semimetals were reported in Refs. \cite{FQHE_Weyl_noB_PRL,FQHE_Weyl_noB_PRB,FQHE_Weyl_3d_PRB}, where the interactions opened a bandgap in the single-particle energy spectrum under a special population of the 
energy levels in such 3D systems. The bandgaps are opened either in the magnetic 3D Weyl semimetals\cite{FQHE_Weyl_noB_PRL,FQHE_Weyl_noB_PRB} or in 3D Weyl semimetals placed in an external magnetic field\cite{FQHE_Weyl_3d_PRB}. Under these conditions the FQHE with fractional quantization of the Hall conductance can be realized in these Weyl semimetals. The fundamental difference between these results and those of ours is that our system is purely two dimensional, which results in the formation of discrete single-particle Landau level spectrum, while in Refs. \cite{FQHE_Weyl_noB_PRL,FQHE_Weyl_noB_PRB,FQHE_Weyl_3d_PRB} the Weyl systems are three dimensional with continuous single-particle energy spectra.

There are multitude of possible reasons behind the selective enhancement of electron-electron interactions at some Landau levels of the system that is discussed above. One of the most likely scenarios is the anisotropy of the system, which in Weyl semimetal systems can sometimes result in some unusual reported effects, such as  the chiral anomaly, which is associated with a variety of 
unusual anisotropic magnetoresistance\cite{anosotropic_magnetoresistance,Annual_review_semimetals}, and a change in the 
nature of the density of states etc. \cite{thermopower_anisotropic_Weyl}. In our case, the anisotropy present in the bismuthene monolayer results in Landau levels that are the mixtures of conventional two-dimensional Landau levels. Such a mixture changes the effective electron-electron interactions within a given Landau level and for some of them, it actually enhances the interactions between electrons compared to the conventional case. For conventional 2D systems with parabolic energy dispersion, 
the anisotropy in the mass term results in suppression of the electron-electron 
interactions\cite{band_mass_anisotropy_conventional_2012}. 
For 2D Weyl semimetal systems discussed above, the anisotropy is of different type. In fact, there are two sources of anisotropy: the tilting term $\Delta$ and the anisotropy of the Fermi velocity. Our analysis shows that the enhancement of electron-electron interactions is mainly due to the tilting term, while the anisotropy of the Fermi velocity has a weak effect on the electron-electron interaction strength and results in its small suppression. 
The spin-orbit coupling has also a subtle role to play in the enhancement of the electron-electron interaction. The unusual behavior of different Landau levels  associated with the energy gaps that we have found here might  
be a signature of the 2D Weyl fermions considered here. From a broader perspective, our present study indicates the possible route to investigate the interaction effects in low-dimensional topological materials, not least the 2D Weyl fermions, by the application of a quantizing magnetic field.

%%%%%%%%%
%1%and Fig.~\ref{fig1}.%

%%%%%

%%%%%%%%%
%1%and Fig.~\ref{fig1}.%

%%%%%

%%%%%%%%%
%1%and Fig.~\ref{fig1}.%

%%%%%

\begin{acknowledgments}
Major funding was provided by Grant No. DE-FG02-01ER15213
from the Chemical Sciences, Biosciences, and Geosciences
Division, Office of Basic Energy Sciences, Office of Science,
US Department of Energy.
Numerical simulations were performed using support by Grant No. DE-SC0007043
from the Materials Sciences and Engineering Division of
the Office of the Basic Energy Sciences, Office of Science,
US Department of Energy, and was also supported in part by the High Performance Computing Center of Central South University. 
\end{acknowledgments}

%%%%%%%%%%%%%%%

%\bibliography{references}

\end{document}